\documentclass[aps,twocolumn,showpacs,pra,superscriptaddress,floatfix]{revtex4}

\usepackage{amsmath,amsthm,amsfonts,amssymb,times,bbm,graphicx}

\usepackage{color}
\usepackage{tabularx}
\usepackage{epsfig}
\usepackage{mathtools}
\usepackage{amsmath}
\DeclareMathOperator{\arcsinh}{arcsinh}
\usepackage{amssymb}
\usepackage{graphicx}
\usepackage{wasysym}
\usepackage{dcolumn}
\usepackage{bm}
\usepackage{epstopdf}
\usepackage{subfigure}
\usepackage{psfrag}
\usepackage{times}
\usepackage{bbm}

\begin{document}

\title{Natural Orbitals and Occupation Numbers for Harmonium: Fermions vs. Bosons}

\author{Christian Schilling}
\affiliation{Institute for Theoretical Physics, ETH Z\"urich, Wolfgang--Pauli--Strasse 27, CH-8093 Z\"urich, Switzerland}


\date{\today}


\begin{abstract} For a quantum system of $N$ identical, harmonically interacting particles in a one-dimensional harmonic trap we calculate for the bosonic and fermionic ground state the corresponding $1$-particle reduced density operator $\hat{\rho}_1$ analytically. In case of bosons $\hat{\rho}_1$ is a Gibbs state for an effective harmonic oscillator. Hence the natural orbitals are Hermite functions and their occupation numbers obey a Boltzmann distribution. Intriguingly, for fermions with not too large couplings the natural orbitals coincide up to just a very small error with the bosonic ones. In case of strong coupling this still holds qualitatively. Moreover, the decay of the decreasingly ordered fermionic natural occupation numbers is given by the bosonic one, but modified by an algebraic prefactor. Significant differences to bosons occur only for the largest occupation numbers. After all the ``discontinuity'' at the ``Fermi level'' decreases with increasing coupling strength but remains well pronounced even for strong interaction.
\end{abstract}


\maketitle

\section{Introduction}\label{sec:intro}
For most quantum systems of $N$ interacting particles it is impossible to solve the corresponding time independent Schr\"odinger equation in the $N$-particle configuration space $\mathcal{C}_N = \{(\vec{x}_1,\ldots,\vec{x}_N)\}$ and determine analytically the eigenenergies $E$ and the eigenfunctions $\Psi$.
As a consequence one typically resorts to approximations, like Hartree-Fock approximation, or as frequently be done in Quantum Chemistry to numerical methods to gain  insight into the quantum system.

For the ground state of identical fermions in an external potential $v(\vec{x})$ it has been proven by Hohenberg and Kohn \cite{Hohenberg} that its energy $E_0$ and (spatial) $1$-particle density,
\begin{equation}\label{density}
n(\vec{x}) = \int \prod_{k=2}^N\mathrm{d}\vec{x}_k\, \rho(\vec{x},\vec{x}_2,\ldots,\vec{x}_N)\,,
\end{equation}
with $\rho(\vec{x}_1,\ldots,\vec{x}_N)=|\Psi(\vec{x}_1,\ldots,\vec{x}_N)|^2$ the probability density on $\mathcal{C}_N$, can also be obtained as the minimizer of an appropriate density functional $E_{v}[n(\cdot)]$. In that case the computation of the ground state properties is effectively reduced to the three-dimensional space, which is much more appealing than to deal with the high-dimensional
$N$-fermion configuration space. However, the functional $E_{v}[\cdot]$ is not known exactly.

The spatial density $n(\vec{x})$ is a special case of the spatial $n$-particle reduced density operator ($n$-RDO)
\begin{eqnarray}\label{n-RDO}
\lefteqn{\rho_n(\vec{x}_1,\ldots,\vec{x}_n,\vec{y}_1,\ldots,\vec{y}_n)}&&\nonumber \\
 &=& \int \prod_{k=n+1}^N \!\mathrm{d}\vec{x}_k\, \Psi(\vec{x}_1,\ldots,\vec{x}_n,\vec{x}_{n+1},\ldots,\vec{x}_N)^\ast \nonumber \\
    && \cdot \Psi(\vec{y}_1,\ldots,\vec{y}_n,\vec{x}_{n+1},\ldots,\vec{x}_N)\,
\end{eqnarray}
$n=1,2,\ldots$. In case of a quantum system with only $1$-body and $2$-body interaction the expectation value of the Hamiltonian $H$ in state $\Psi(\vec{x}_1,\ldots,\vec{x}_N)$ can be expressed just by $\rho_1$ and $\rho_2$. This is the reason why reduced density operators in coordinate and coordinate-spin space have been studied intensively. For reviews the reader is referred to \cite{Col2,Dav}. Not that much is known yet about the set of possible $2$-particle density operators $\rho_2$, which do arise via Eq.\ (\ref{n-RDO}) from pure (antisymmetric) states. In particular it is known that the task to find this set is QMA-complete \cite{Liu2007}, expressing the high complexity of that problem. Therefore, the major activity has been concentrated onto the study of $\rho_1$ and its eigenvalue equation
\begin{equation}
\int\!\mathrm{d}y^3\,\rho_1(\vec{x},\vec{y}) \,\chi_k(\vec{y}) = \lambda_k\, \chi_k(\vec{x})\,,
\end{equation}
$k=1,2,3\ldots$. Its eigenfunctions $\chi_k(\vec{x})$ are called \emph{natural orbitals} (or if spin is also taken into account natural spin orbitals) and its eigenvalues $\lambda_k$, the occupation numbers of those orbitals, \emph{natural occupation numbers}. The physical relevance of those states and occupation numbers has been discussed (see e.g.\ Refs. \cite{Dav} and \cite{Helbig} and references therein). Usually, the $\lambda_k$'s are ordered, $\lambda_1 \geq \lambda_2\geq \ldots$, and normalized to the particle number,
\begin{equation}
\int\!\mathrm{d}x^3\,\rho_1(\vec{x},\vec{x}) = \int\!\mathrm{d}x^3\, n(\vec{x}) =\sum_{k=1}^\infty \lambda_k = N\,.
\end{equation}
In the case of identical fermions the Pauli exclusion principle, $1\geq \lambda_k\geq 0$, significantly restricts the occupation numbers. However, the antisymmetry of the $N$-fermion wave function is even a stronger restriction and its influence on natural occupation numbers amounts to the fermionic $1$-body quantum marginal problem, which asks whether given natural occupation numbers can arise from an antisymmetric $N$-particle state.
In a ground breaking work this problem was solved by A.Klyachko \cite{Kly2,Kly3} and it was shown that the antisymmetry implies further restrictions on the occupation numbers, the so-called generalized Pauli constraints. Recently, we have provided the first analytical evidence of their physical relevance for ground states \cite{CS2013}: For the $3$-Harmonium system, by anticipating some results that we will prove in the present work, we have shown that the generalized Pauli constraints for not too strong interaction are surprisingly well-saturated by the natural occupation numbers. This behavior of the natural occupation numbers to lie very close to the boundary of the allowed region is called quasi-pinning and has an important physical relevance, since it implies that the structure of the corresponding $N$-fermion states is significantly simpler (contains contributions from just a few Slater determinants) \cite{Kly1, CS2013}. Further results on the quantum marginal problem can be found e.g.\ in \cite{Daftuar, Brav, Borl1972, MC, Rus2, Higuchi, Kly4, EisertGaussian}.

In the present paper we will not study the quantum marginal problem, but investigate the properties of natural orbitals and the natural occupation numbers for the so-called ``$N$-Harmonium'', a model of $N$ identical particles in a one-dimensional harmonic trap, which are coupled harmonically to each other. A realization of this model might be an ultracold gas of particles in a harmonic trap, where the Coulomb interaction is replaced by a harmonic one. But it can also be interpreted as a harmonic lattice where the trap potential acts as an on-site potential for each atomic displacement from its equilibrium position.

For that model one can calculate its eigenfunctions exactly and also its $1$-RDO for arbitrary $N$. This can be done for both particle types, spinless bosons and spinless fermions. It is our major goal to investigate possible similarities between the bosonic and fermionic natural orbitals and natural occupation numbers.

Harmonic systems were already studied before. For the Harmonium with \emph{two} spinless bosons in one dimension $\rho_1$ was calculated for the ground state \cite{Rob}. The same was done for \emph{two} electrons in three dimensions \cite{Dav}. $\rho_1$ was also derived for the ground state of a harmonic chain of spinless bosons with nearest neighbor coupling and an external harmonic potential \cite{Peschel1999}. For these three different harmonic models $\rho_1(x,y)$ is an exponential function with an exponent bilinear in $x$ and $y$. Therefore, $\rho_1(x,y)$ can be represented as a Gibbs state of an effective harmonic oscillator. These findings are not surprising, since the ground state of \emph{spinless} bosons with arbitrary harmonic interactions, e.g.\ in one dimension, is an exponential function bilinear in the particle coordinates. Accordingly, the $n$-RDO $\rho_n$ from Eq.\ (\ref{n-RDO}) keeps this form for all $n$. For fermions with harmonic interactions the result of Ref. \cite{Dav} for two electrons in its singlet ground state seems to be the only one for the $1$-RDO $\rho_1$. $\rho_1$ has been calculated for free spinless fermions \cite{Peschel2003}. The corresponding Hamiltonian in second quantized form is bilinear in the fermionic creation and annihilation operators. Its eigenvalue problem is solved by diagonalizing the bilinear form. Then, it was shown that $\rho_1$ again can be represented by a Gibbs state with an effective quadratic Hamiltonian \cite{Peschel2003}. However, such free fermionic Hamiltonians are different to Hamiltonians for fermions with harmonic interactions. Free bosonic (see e.g.\ \cite{EisertHarmChain,EisertHarmLattice,Eisert2007}) and free fermionic systems (see e.g.\ \cite{Wolf2006,Eisert2007}) were also investigated using concepts from quantum information theory with focus on entropy and entanglement. This involves the reduced density operator for, e.g.\ a bipartite systems. However, these concepts are not the issue of our present contribution.

The outline of our paper is as follows. In the next section, we describe details of the ``$N$-Harmonium'' and its eigenfunctions and derive the $1$-RDO for the case of identical spinless bosons and identical spinless fermions in their ground state. In Sec. \ref{sec:NO} we analytically calculate for bosons (b) and fermions (f), respectively, the natural occupation numbers $\lambda_k^{(\alpha)}$ and natural orbitals $\chi_k^{(\alpha)}$ , $\alpha= b,f$. Since for fermions this is only feasible for the regime $k \gg 1$, we also present some numerical results for arbitrary $k$. The final section, Sec. \ref{sec:Summary}, contains a summary and discussions of the results. Technical details are presented in the appendices.

\section{Model and $1$-Particle Reduced Density Operator}\label{sec:model}
In this section we introduce the ``$N$-Harmonium'' and describe how its eigenvalue problem can be solved. It is demonstrated how the corresponding $1$-RDO $\rho_1(x,y)$ for the bosonic and fermionic ground state can be calculated analytically.
\subsection{Model and its eigenfunctions}
We consider a system of $N$ (spinless) identical particles with mass $m$ and scalar coordinates $x_i, i=1,2,\ldots,N$. The Hamiltonian is given by
\begin{equation}\label{hamiltonianX}
H_N^{(X)} = \sum_{i=1}^{N} \left(\frac{p_i^2}{2 m} +\frac{1}{2} m \omega^2 {x_i}^2\right) + \frac{1}{2} D \sum_{1\leq i<j\leq N} (x_i-x_j)^2.
\end{equation}
The particles feel an external harmonic potential $\frac{1}{2} m \omega^2 {x}^2$ and interact harmonically with a coupling constant $D$, which may be attractive ($D>0$) or  repulsive ($D<0$). For the repulsive regime we require $-\frac{m \omega^2}{N}<D$ to guarantee the existence of bound states.
The potential term in Eq.\ (\ref{hamiltonianX}) can be expressed as $\frac{1}{2} x_i \mathcal{D}_{ij} x_j$ (here and in the following summation convention is used), with
\begin{equation}
\mathcal{D}_{ij} =(m \omega^2 +(N-1)D)\, \delta_{ij}-D(1-\delta_{ij}).
\end{equation}
In shorthand notation $\vec{x}=(x_1,\ldots,x_N)^T$ and $\vec{p}=(p_1,\ldots,p_N)^T$ the Hamiltonian reads
\begin{equation}
H_N^{(X)} = \frac{\vec{p}^{\,2}}{2m} + \frac{1}{2}\vec{x}^T \mathcal{D} \vec{x}\,.
\end{equation}
The real and symmetric matrix $\mathcal{D}=(\mathcal{D}_{ij})$ can easily be diagonalized by a $N$-dimensional orthogonal matrix $S = (\vec{e}_1,\vec{e}_2,\ldots,\vec{e}_N)^T$ with orthonormalized column vectors
\begin{eqnarray}\label{coordtrafo}
\vec{e}_1 &=& \frac{1}{\sqrt{N}}(1,\ldots,1)^T \nonumber \\
\vec{e}_k & = & \frac{1}{\sqrt{k(k-1)}}(\underbrace{1,\ldots,1}_{k-1},-1,0,\ldots,0)^{T} \,\,,k\geq2\,.
\end{eqnarray}
It follows
\begin{equation}\label{coordtrafoortho}
S \mathcal{D} S^T = \mathcal{D}_0 \,\,\,,(\mathcal{D}_{0})_{ij} = d_i \delta_{ij},
\end{equation}
where $d_1 \equiv d_{-}:= m \omega^2$ and $d_k \equiv d_{+}:= m \omega^2+N D$ for $k\geq 2$. Hence, the coordinate transformation
\begin{equation}
y_i = S_{ij} x_j
\end{equation}
decouples the $N$ coordinates and the Hamiltonian in the new coordinates $y_i$ and corresponding momenta $\pi_i$ reads
\begin{equation}\label{hamiltonianX2}
H_N^{(Y)} = \sum_{i=1}^{N} \left(\frac{\pi_i^2}{2 m} +\frac{1}{2} m \omega_i^2 {y_i}^2\right)
\end{equation}
with harmonic oscillator frequencies $\omega_{j}=\sqrt{d_j/m} $, $j=1,2,\ldots,N$. Since $y_1= \frac{x_1+\ldots+x_N}{\sqrt{N}}$
the oscillator with index $i=1$ describes the center of mass motion in the harmonic trap. Clearly, the corresponding frequency $\omega_1 \equiv \omega_{-}:= \omega$ is not affected by the interaction between the $N$ particles. The remaining $N-1$ harmonic oscillators in Eq.\ (\ref{hamiltonianX2}) describe the relative motion, all with the same frequency $\omega_k \equiv \omega_{+}:= \sqrt{\omega^2+\frac{N D}{m}}, k=2,\ldots,N$.
Note that the decoupling of the $N$ coordinates can also be obtained by  use of the Jacobian coordinates \cite{harmOsc2012}.

The spectrum of Hamiltonian (\ref{hamiltonianX2}) is well-known. The eigenenergies are given by
\begin{equation}
E_{\bm{\nu}} = \hbar \omega_{-}(\nu_1 + \frac{1}{2}) + \hbar \omega_+ \sum_{i=2}^{N} (\nu_i+ \frac{1}{2}) \,,
\end{equation}
$\bm{\nu}\equiv (\nu_1,\ldots,\nu_N)  , \nu_{i}=0,1,2,...$. We introduce the $\nu$-th Hermite function, an eigenfunction of a single $1$-dimensional harmonic oscillator with natural length scale $l$,
\begin{equation}\label{Hermitefunction}
\varphi_{\nu}^{(l)}(y) = \pi^{-\frac{1}{4}} l^{-\frac{1}{2}} (2^{\nu} {\nu}!)^{-\frac{1}{2}} H_{\nu}(\frac{y}{l}) \mbox{e}^{-\frac{y^2}{2 l^2}}\,
\end{equation}
where $H_{\nu}$ is the $\nu$-th Hermite polynomial. Then,  using the shorthand  notation $\varphi_{\nu}^{(+/-)}(y) \equiv \varphi_{\nu}^{(l_+/{l_-})}(y)$ the  eigenfunctions of the Hamiltonian (\ref{hamiltonianX2}) read
\begin{equation}
\Psi_{\bm{\nu}}(\vec{y}) = \varphi_{\nu_1}^{(-)}(y_1) \prod_{i=2}^N \varphi_{\nu_i}^{(+)}(y_i)\,.
\end{equation}
The corresponding natural length scales $l_{-}$ and $l_{+}$  are given by $l_j= \sqrt{\hbar/\sqrt{m d_j}}$ and are related to the coupling constants $D$ and $m \omega^2$ by
\begin{equation}\label{LvsHook}
\frac{N D}{m \omega^2} = \left(\frac{l_-}{l_+}\right)^4-1\,.
\end{equation}
For macroscopic particle numbers one should rescale $D$ by N, i.e.\ $D \rightarrow D/N$, in order that the energy per particle is of order one in $N$.
So far, these eigenfunctions do not describe bosonic or fermionic particles, since the required symmetry for the wave function under particle exchange is not given yet. Since we will study bosons and fermions we need to restrict Eq.\ (\ref{hamiltonianX}) to the $N$-boson Hilbert space of symmetric wave functions and the $N$-fermion Hilbert space of antisymmetric wave functions, respectively. In the following we will focus onto the ground states for both particle types.

The ground state $\Psi_{0}^{(b)}$ for spinless \emph{bosons} coincides with the absolute $N$-particle ground state, i.e.\ it is characterized by  $\nu_{i}=0$ for $i=1,2,...,N$.
Moreover, by using the orthogonal character of the transformation matrix $S$ (cf. Eqs.\ (\ref{coordtrafo}), (\ref{coordtrafoortho})) and by reintroducing the physical coordinates $x_i$ we find
\begin{eqnarray}\label{gsbosons}
\Psi_{0}^{(b)}(\vec{x})& =& \mathcal{N}  \exp{\left[-\frac{y_1(\vec{x})^2}{2 {l_{-}}^2} -\frac{1}{2 {l_{+}}^2} \sum_{k=2}^N y_k(\vec{x})^2\right]} \nonumber \\
&=& \mathcal{N}  e^{- A \vec{x}^2 + B_N (x_1+\ldots+x_N)^2} \,,
\end{eqnarray}
where $\mathcal{N}$ is the normalization factor and
\begin{equation}\label{parameterAB}
A \equiv \frac{1}{2 {l_{+}}^2}    \,\,\,,\,B_N\equiv \frac{1}{2}\left(\frac{1}{{l_{+}}^2} - \frac{1}{{l_{-}}^2}\right)\,.
\end{equation}
Note that for zero interaction, $B_N$ vanishes, since $l_-=l_+$.

For spinless \emph{fermions} the ground state $\Psi_{0}^{(f)}$ can be found by applying the antisymmetrizing operator to the states $\Psi_{\bm{\nu}}(\vec{y}(\vec{x}))$ with $\nu_{i}=i-1$ for  $i=1,2,...,N$. This was done in Ref. \cite{harmOsc2012} and one finds
\begin{equation}\label{gsfermions}
\Psi_{0}^{(f)}(\vec{x}) = \mathcal{N}\, \left[\prod_{1\leq i<j\leq N}(x_i-x_j)\right] \,e^{- A \vec{x}^2 + B_N (x_1+\ldots+x_N)^2} \,.
\end{equation}
The exponent in Eq.\ (\ref{gsfermions}) is the same one as for the bosonic counterpart, Eq.\ (\ref{gsbosons}), and is after all symmetric under particle exchange. In particular, this means that all the differences between fermions and bosons are arising just from the antisymmetric polynomial in front of the exponential function, the Vandermonde determinant,
\begin{equation}
\prod_{1\leq i<j\leq N}(x_i-x_j) = \left|\begin{array}{lll}\,1&\ldots&\,1\\x_1&\ldots&x_N\\ \,\vdots& &\,\vdots\\ x_1^{N-1}&\ldots& x_N^{N-1} \end{array}\right|\,.
\end{equation}

\subsection{$1$-Particle Reduced Density Operator}
The calculation of the $1$-RDO $\rho_1^{(b)}(x,y)$ for the \emph{bosonic} ground state is straightforward for arbitrary particle number $N$ (see Appendix \ref{sec:appBosons}). One gets
\begin{equation}\label{1RDOb}
\rho_1^{(b)}(x,y) = c_N\, \exp{\left[-a_N (x^2+y^2) +b_N x y\right]}
\end{equation}
with (recall Eq.\ (\ref{parameterAB}))
\begin{eqnarray}\label{parameterab}
b_N&=&\frac{(N-1)B_N^2}{A-(N-1)B_N}\,\,,\,a_N=(A-B_N)-\frac{1}{2}b_N \nonumber \\
c_N&=&N \sqrt{\frac{2 a_N-b_N}{\pi}}\,.
\end{eqnarray}
Note that $\rho_1^{(b)}$ is normalized to the particle number $N$, i.e.\ $\int \!\mathrm{d}x \,\rho_1^{(b)}(x,x)=N$\,.
This result resembles those in Refs. \cite{Rob,Dav,Peschel1999}. The difference to Ref. \cite{Peschel1999} is that the coefficients $a_N, b_N$ of the bilinear exponent can be expressed explicitly by both length scales $l_-, l_+$ for \emph{all} $N$ (cf. Eq.\ (\ref{parameterAB}) and (\ref{parameterab})).

For \emph{fermions}, the explicit computation of $\rho_1^{(f)}$ for arbitrary $N$ is much more involved. Again, as for the $N$-particle ground states, the exponential part of the fermionic $1$-RDO coincides with the bosonic one. The Vandermonde determinant in front of the exponential term in Eq.\ (\ref{gsfermions}) leads to an additional symmetric polynomial $F_N(x,y)$ of degree $2(N-1)$ and with only even order monomials (see Appendix \ref{sec:appFermions}):
\begin{equation}
F_N(x,y) = \sum_{\nu=0}^{N-1} \sum_{\mu=0}^{2 \nu} \,c_{\nu,\mu}\, x^{2\nu-\mu} y^{\mu}\,.
\end{equation}
The coefficients $c_{\nu,\mu}$ depend on the model parameters and fulfill $c_{\nu,\mu} = c_{\nu,2\nu-\mu}$. Accordingly, we have
\begin{equation}\label{1RDOf}
\rho_1^{(f)}(x,y) = F_N(x,y)\, \exp{\left[-a_N (x^2+y^2) +b_N x y\right]},
\end{equation}
which is again normalized to $N$. The expression for the coefficients $c_{\nu,\mu}$  is rather cumbersome (see Eq.\ (\ref{1RDOfUint})). The number of terms contributing to $c_{\nu,\mu}$ increases with increasing $N$.
As an example we present the explicit result for $N=3$:
\begin{eqnarray}
F_3(x,y) &=& d_3 \big[ C_1 (x^4+y^4)+C_2 (x^3 y+x y^3)+C_3 x^2 y^2\nonumber \\
&& C_4 (x^2+y^2) +C_5 x y + C_6
\end{eqnarray}
with
\begin{eqnarray}
C_1&=&\frac{1}{24} \big(96 A^4 B_N^2-480 A^3 B_N^3+600 A^2 B_N^4\big) \nonumber \\
C_2&=&\frac{1}{6} \big(-96 A^5 B_N+720 A^4 B_N^2-1824 A^3 B_N^3 \nonumber \\
&=&+1560 A^2 B_N^4\big)\nonumber
\end{eqnarray}
\begin{eqnarray}
C_3&=&\frac{1}{4} \big(64 A^6-640 A^5 B_N+2464 A^4 B_N^2-4320 A^3 B_N^3 \nonumber\\
&&+2904 A^2 B_N^4\big)\nonumber \\
C_4&=&\frac{1}{2} \big(-8 A^5+72 A^4 B_N-264 A^3 B_N^2+460 A^2 B_N^3 \nonumber \\
&&-312 A B_N^4\big)\nonumber \\
C_5&=&8 A^5-48 A^4 B_N+72 A^3 B_N^2+44 A^2 B_N^3-120 A B_N^4\nonumber \\
C_6&=&3 A^4-24 A^3 B_N+75 A^2 B_N^2-108 A B_N^3+60 B_N^4\nonumber \\
d_3&=& \frac{\sqrt{A^2-3 A B_N}}{\sqrt{2 \pi } \left(A-2 B_N\right){}^{9/2}}\,.
\end{eqnarray}

\section{Natural Orbitals and their Occupation Numbers}\label{sec:NO}

In this section we will discuss the eigenvalue equation for the bosonic and fermionic $1$-RDO. For a finite but arbitrary number of bosons we can determine exactly the  natural occupation numbers $\lambda_k^{(b)}$ and natural orbitals $\chi_k^{(b)}$, and for fermions this can only be done for  $k$ sufficiently large.

\subsection{Bosons}\label{sec:Bosons}
The $1$-RDO $\rho_1^{(b)}$ for bosons, Eq.\ (\ref{1RDOb}), has the form of a Gibbs state in coordinate representation \cite{feyn}
\begin{widetext}
\begin{eqnarray}\label{Gibbs}
\rho_1^{(b)}(x,y)&=& \frac{1}{Z_{eff}}\, \langle x|\exp{[-\beta_N \hat{H}_{eff}]}|y\rangle\nonumber \\
&=& N \sqrt{\frac{1}{\pi L_N^2}\tanh{(\beta_N \hbar \Omega_N/2)}}\,\exp{\left(-\frac{1}{2 L_N^2 \sinh{(\beta_N\hbar \Omega_N)}}\left[(x^2+y^2)\cosh{(\beta_N\hbar \Omega_N)}- 2 x y\right]\right)}
\end{eqnarray}
\end{widetext}
where  $\hat{H}_{eff}$ is the effective Hamiltonian for a single harmonic oscillator with mass $M_N$, frequency $\Omega_N$ and length scale $L_N = \sqrt{\frac{\hbar}{M_N  \Omega_N}}$:
\begin{equation}\label{Hamiltonianeff}
\hat{H}_{eff} = \frac{1}{2} \hbar \Omega_N \left[-L_N^2\frac{\mathrm{d}^2}{\mathrm{d}x^2}+\frac{1}{L_N^2} x^2\right]\,.
\end{equation}
From Eqs.\ (\ref{1RDOb}), (\ref{Gibbs}) and $\rho_1^{(b)}(x,y) = \langle x| \hat{\rho}_1^{(b)}|y\rangle$ we obtain
\begin{equation}\label{GibbsOp}
\hat{\rho}_1^{(b)} = \frac{1}{Z_{eff}}\,\exp{[-\beta_N \hat{H}_{eff}]}\,,
\end{equation}
with
\begin{eqnarray}\label{effectiveoscill}
L_N &=& (4 a_N^2-b_N^2)^{-\frac{1}{4}} \nonumber \\
\beta_N \hbar \Omega_N &=& \arcsinh \left(\frac{1}{L_N^2 b_N}\right) \nonumber \\
Z_{eff}&=& \frac{N}{2}\,\left[ \sinh{\left(\frac{\beta_N \hbar \Omega_N}{2}\right)}\right]^{-1}\,.
\end{eqnarray}

\noindent These quantities can also be expressed by the original parameters $l_-$ and $l_+$, only:
\begin{widetext}
\begin{equation}
L_N = \sqrt{l_- l_+}\,\left[\frac{(N-1)l_+^2 +l_-^2}{l_+^2 +(N-1)l_-^2}\right]^{\frac{1}{4}} \,\,\,,\, \beta_N \hbar \Omega_N = \arcsinh{ \left[\frac{2 l_+ l_- \sqrt{[(N-1)l_+^2+l_-^2][l_+^2+(N-1)l_-^2]}}{(1-1/N)(l_+^2-l_-^2)^2}\right]}\,.
\end{equation}
\end{widetext}
Note that $L_N\rightarrow (N-1)^{-\frac{1}{4}}\,l_-\,\left(l_+/l_-\right)^{\frac{1}{2}}$, $\beta_N \hbar \Omega_N\rightarrow \left(2N/\sqrt{N-1}\right) \,l_+/l_-$ for $l_+/l_-\rightarrow 0$ corresponding to $D\rightarrow \infty$, and $L_N\rightarrow l_-$, $\beta_N \hbar \Omega_N\rightarrow  \infty$ for $l_+/l_-\rightarrow 1$, i.e.\ $D \rightarrow 0$. The result (\ref{GibbsOp}) demonstrates that the $1$-RDO can exactly be represented by the Gibbs state of an effective harmonic oscillator at a ``temperature'' $T_N = 1/(k_B \beta_N)$. That $\hat{\rho}_1$ is a Gibbs state for an effective harmonic oscillator has already been shown in \cite{Peschel1999} for a harmonic chain with nearest neighbor interactions. Due to the permutation invariance of the harmonic potential of our model, the parameters of the effective Hamiltonian can be calculated explicitly as functions of $l_-$ and $l_+$ (see Eqs.\ (\ref{parameterAB}), (\ref{parameterab}), (\ref{Hamiltonianeff}) and (\ref{effectiveoscill})).

For $D=0$, i.e.\ non-interacting bosons, it follows $l_-=l_+$. For that case, the ``temperature'' is zero.
For $\hat{\rho}_1^{(b)}$ from Eq.\ (\ref{GibbsOp}) it is easy to determine the natural orbitals $\chi_k^{(b)}(x)$ and the corresponding occupation numbers $\lambda_k^{(b)}$, which obey the eigenvalue equation
\begin{equation}
\hat{\rho}_1^{(b)} \chi_k^{(b)} = \lambda_k^{(b)} \chi_k^{(b)}\,.
\end{equation}
By recalling the Hermite functions $\varphi_k^{(l)}(x)$ (see Eq.\ (\ref{Hermitefunction})) we find $\chi_k^{(b)}(x) = \varphi_k^{(L_N)}(x)$.
Moreover, the natural occupation numbers obey the Boltzmann law
\begin{equation}\label{NONb}
\lambda_k^{(b)} = N \left[1-\exp{(-\beta_N \hbar \Omega_N)}\right]\, e^{-(\beta_N \hbar \Omega_N)k}\,,k=0,1,\ldots
\end{equation}
It is obvious that  $\lambda_k^{(b)}$ fulfill the standard normalization $\sum_{k=0}^{\infty} \lambda_k^{(b)} = N$.

\subsection{Fermions: Analytical Results}\label{sec:FermionsAnalytical}
Although it seems to be impossible to solve analytically the eigenvalue problem for the fermionic $1$-RDO $\rho_1^{(f)}(x,y)$  for arbitrary $N$ we calculate in the following most of the main features of the natural occupation numbers and natural orbitals.

Since $\rho_1^{(f)}(x,y)$ has the same exponential factor as $\rho_1^{(b)}(x,y)$ it is reasonable to expand $\chi^{(f)}$ w.r.t. the bosonic natural orbitals $\chi_m^{(b)}(x)$, the Hermite functions $\varphi_m^{(L_N)}(x) \equiv \langle x | m\rangle$, i.e.
\begin{equation}
|\chi^{(f)}\rangle = \sum_{m=0}^{\infty} \zeta_m |m\rangle\,.
\end{equation}
The eigenvalue equation for $\rho_1^{(f)}(x,y)$ reduces to a discrete equation for the expansion coefficients $\{\zeta_m\}$,
\begin{equation}\label{eigenprobmatrixf}
\sum_{n=0}^{\infty} \langle m |\hat{\rho}_1^{(f)} | n \rangle \zeta_n = \lambda^{(f)} \zeta_m\,.
\end{equation}
In the following we choose the particle number $N$ arbitrary, but fixed. Using Eq.\ (\ref{coefrelNOf}) from the Appendix \ref{sec:appEigenvalue}, Eq.\ (\ref{eigenprobmatrixf}) for sufficiently large $m$ reduces to
\begin{equation}\label{coefrelation}
m^{N-1} e^{-\beta_N \hbar  \Omega_N (m+\frac{1}{2})}\,\sum_{r=-(N-1)}^{N-1}\,h_{m,m-2r}\,\zeta_{m-2r} \simeq \lambda^{(f)} \zeta_m\,.
\end{equation}
Here we used, e.g.\ $\sqrt{m+r} \simeq \sqrt{m}$  for $m \gg 1$  and $r=O(1)$.
For illustration, we discuss this equation for  $N=2$ (for larger $N$ one can proceed similarly), i.e.
\begin{equation}\label{coefrelation2}
m e^{-\beta_2 \hbar  \Omega_2 (m+\frac{1}{2})}\left[h_- \zeta_{m-2} + h_0 \zeta_m +h_+ \zeta_{m+2}\right] \simeq \lambda^{(f)} \zeta_m,
\end{equation}
where $h_0 \equiv h_{m,m}, h_{\pm} \equiv h_{m,m\pm 2}$ do not depend on $m$. For vanishing interaction the eigenfunctions of $\rho_1^{(f)}(x,y)$ are the Hermite functions $\varphi_k^{(L_N)}$. Accordingly, we can label the eigenfunctions by $k$ and find for that case $|\chi_k^{(f)}\rangle = |k \rangle$ and thus $\zeta_m^{(k)} = \delta_{k,m}$. Turning on the interaction we expect the main contributions to $|\chi_k^{(f)}\rangle$ coming still from $|k\rangle$. Since $\zeta_{k\pm 2}$ is at most of the same order as $\zeta_k$, we conclude from Eq.\ (\ref{coefrelation2}) with $m=k$ that
\begin{equation}\label{NONf}
\lambda^{(f)} \rightarrow \lambda^{(f)}_k \sim  k \, e^{-\beta_2 \hbar  \Omega_2 (k+\frac{1}{2})}\,, k\gg 1\,.
\end{equation}
For each $k$, $\zeta_m^{(k)}$ for $m \rightarrow \infty$ decays to zero due to the normalization of  $\chi_k^{(f)}$. Therefore, as a consistency ansatz, let us assume that
$|\zeta_m^{(k)}/\zeta_{m-2}^{(k)}| \ll 1$ for $m\gg k$. This together with (\ref{coefrelation2}) and (\ref{NONf}) leads to
\begin{equation}\label{coefdecayr}
\frac{\zeta_m^{(k)}}{\zeta_{m-2}^{(k)}} \sim \frac{m}{k}\, e^{-\frac{1}{4}\beta_2 \hbar  \Omega_2 (m-k)}\,,
\end{equation}
which is indeed consistent with our assumption $|\zeta_m^{(k)}/\zeta_{m-2}^{(k)}| \ll 1$. Moreover, from (\ref{coefdecayr}) we obtain the Gaussian decay behavior
\begin{equation}
\zeta_m^{(k)} \sim  e^{-\frac{1}{4}\beta_2 \hbar  \Omega_2 (m-k)^2}
\end{equation}
for $k\gg1$ and $m\gg k$.
For the opposite regime, $1\ll m \ll k$, and taking $h_0 = O(m^0)$ into account we have
\begin{eqnarray}
|m e^{-\beta_2 \hbar  \Omega_2 (m+\frac{1}{2})} h_0 \zeta_m^{(k)}| & \gg & |\lambda_k \zeta_m^{(k)}| \nonumber \\
&\propto & |k  e^{-\beta_2 \hbar  \Omega_2 (k+\frac{1}{2})} \zeta_m^{(k)}|\,.
\end{eqnarray}
Eq.\ (\ref{coefrelation2}) then implies
\begin{equation}\label{coefrelationl}
\zeta_m^{(k)} + \frac{h_-}{h_0} \,\zeta_{m-2}^{(k)} + \frac{h_+}{h_0} \,\zeta_{m+2}^{(k)}\simeq 0\,,
\end{equation}
which is solved by an exponential
\begin{equation}\label{coefdecayl}
\zeta_m^{(k)} \sim  e^{\alpha_2 (m-k)}\,,
\end{equation}
where $\alpha_2$ depends on $h_0, h_{\pm}$, but not on the orbital index $k$.  $\alpha_2$ can be determined by plugging the ansatz (\ref{coefdecayl}) into Eq.\ (\ref{coefrelationl}) and solving the emerging quadratic equation for $e^{2\alpha_2}$. Since $\zeta_m^{(k)}$ for $1 \ll m \ll k$ should decay with decreasing $m$ the root with $Re(\alpha_2) > 0$ should be taken.

By just repeating all these steps for Eq.\ (\ref{coefrelation}) we find for arbitrary $N$
\begin{equation}\label{NONfN}
\lambda^{(f)} \rightarrow \lambda^{(f)}_k \sim  k^{N-1}  e^{-\beta_N \hbar  \Omega_N (k+\frac{1}{2})}\,,k\gg 1\,.
\end{equation}
The decay behavior of $\zeta_m^{(k)}$  for $m\gg k$ is again Gaussian,
\begin{equation}\label{coefdecayrN}
\zeta_m^{(k)} \sim  e^{-\frac{1}{4(N-1)}\beta_N \hbar  \Omega_N (m-k)^2}\,,
\end{equation}
and exponential for $1\ll m \ll k$ ,
\begin{equation}\label{coefdecaylN}
\zeta_m^{(k)} \sim  e^{\alpha_N (m-k)}\,,
\end{equation}
where $\alpha_N$ depends on $h_0, h_{\pm r} \equiv h_{m,m\pm 2 r}, r=1,\ldots,N-1$, but not on the orbital index $k$. $\alpha_N$  is the root of a polynomial of degree $2(N-1)$ for which $Re(\alpha_N) > 0$.

\subsection{Fermions: Numerical Results}\label{sec:FermionsNumerical}
In order to check the analytical predictions in Sec. \ref{sec:FermionsAnalytical}   for the natural occupation numbers $\lambda_k^{(f)}$ and the natural orbitals $\chi_k^{(f)}$ we have solved Eq.\ (\ref{eigenprobmatrixf}) numerically for $N=3$ and $N=5$, by representing $\hat{\rho}_1^{(f)}$ and the states $|\chi_k^{(f)}\rangle$ again w.r.t to the bosonic natural orbitals $|\chi_m^{(b)}\rangle$ and then truncating the corresponding matrix $((\rho_1^{(f)})_{n,m})$ and vectors $(\zeta_m^{(k)}), \zeta_m^{(k)} \equiv \langle \chi_m^{(b)} |\chi_k^{(f)}\rangle $ at $m_{max}$. All the results presented here are obtained with $m_{max}=500$. As dimensionless interaction strengths we choose $l_+/l_- = 4/5, 1/2$ and $1/3$, which corresponds (according to Eq.\ (\ref{LvsHook})) to $ND/(m \omega^2) = 369/256 \simeq 1.44, 15$ and $80$.
The numerical calculations in particular allow us to investigate $\lambda_k$ for the regime $k = O(1)$.
\begin{figure}[!h]
\includegraphics[width=8cm]{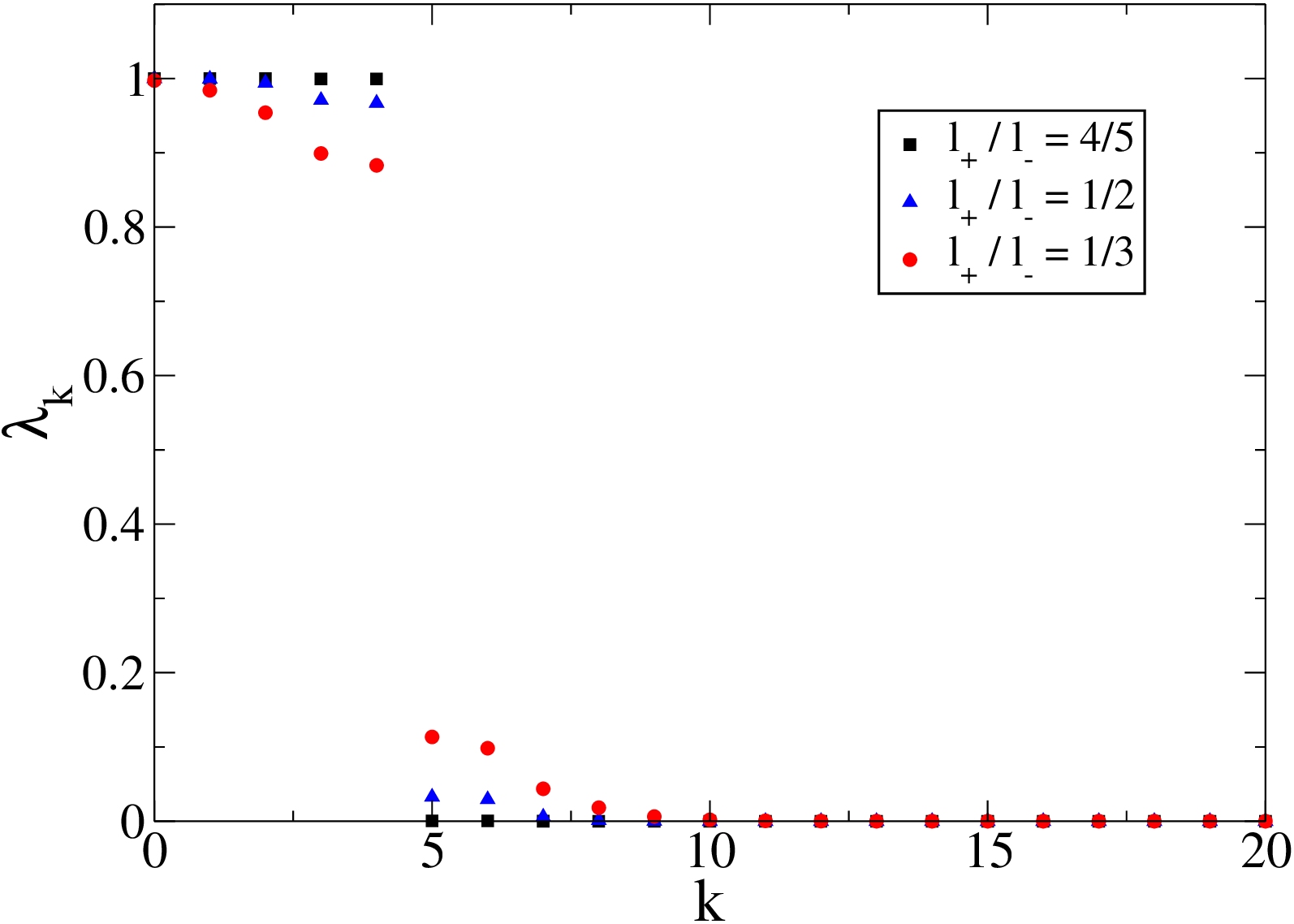}
\centering
\caption{Natural occupation numbers $\lambda_k^{(f)}$ for $N=5$ and three different interaction strengths (see legend).}
\label{fig:non}
\end{figure}
Figure \ref{fig:non} depicts the natural occupation numbers $\lambda_k^{(f)}$ for three different coupling strengths and  $N=5$. Note, even  for quite strong interaction the ``gap'' at the ``Fermi level'' is still well pronounced.
\begin{figure}[!h]
\includegraphics[width=8cm]{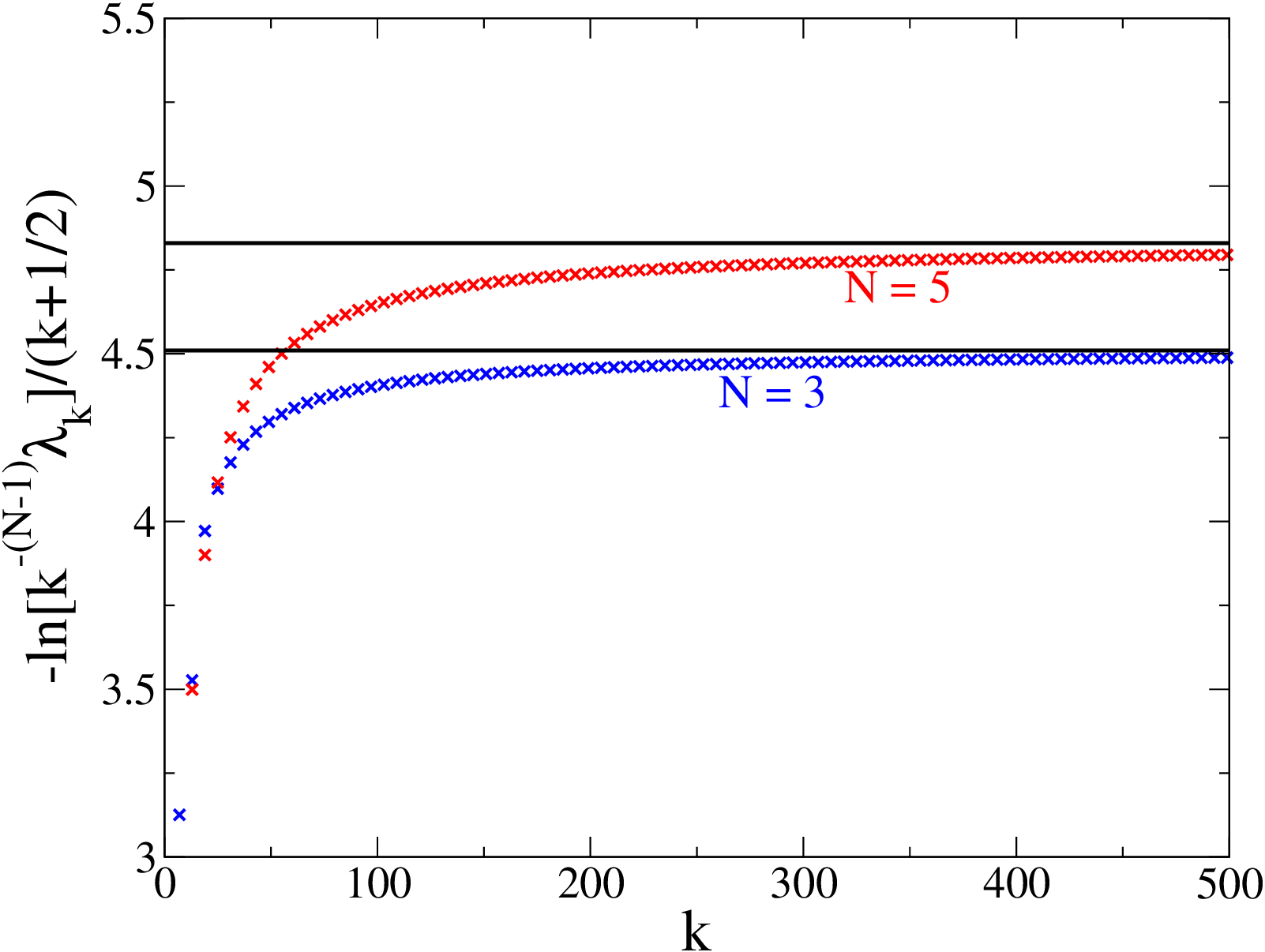}
\centering
\caption{$k$-dependence of $-\ln{(\lambda_k k^{-(N-1)})}/(k+\frac{1}{2})$} for $N=3,5$ and interaction $l_+/l_- = 4/5$. The horizontal lines represent the asymptotic values $\beta_N \hbar \Omega_N$
\label{fig:nonbeta}
\end{figure}

In Figure \ref{fig:nonbeta} we verify the dominant Boltzmann-like behavior found in Eq.\ (\ref{NONfN}) for interaction strength $l_+/l_- = 4/5$ by plotting the $k$-dependence of $-\ln{(\lambda_k k^{-(N-1)})}/(k+\frac{1}{2})$,
which should converge for $k\rightarrow \infty$ to the constant $\beta_N \hbar \Omega_N$. Indeed, this happens since the curves are approaching the values $\beta_3 \hbar \Omega_3 \simeq 4.51$ and $\beta_5 \hbar \Omega_5 \simeq 4.83$ quite well.

\begin{figure}[!h]
\includegraphics[width=8cm]{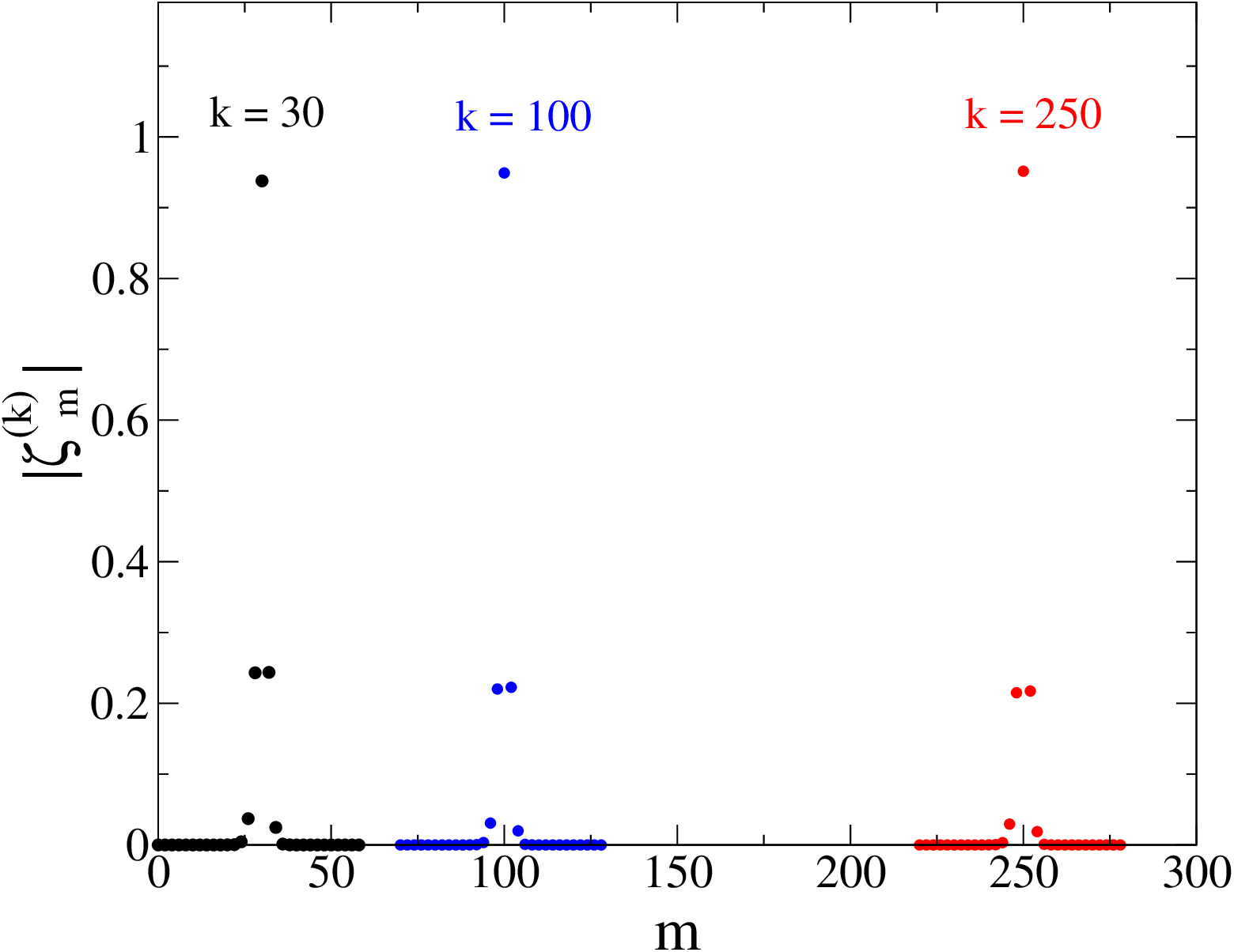}
\centering
\caption{Expansion coefficients $\zeta_m^{(k)} \equiv \langle \chi_{m}^{(b)} | \chi_{k}^{(f)} \rangle$} for the natural orbitals $\chi_{k}^{(f)}$, $k=30,100$ and $250$ for $N=5$ and $l_+/l_- = 1/3$ for the relevant regime $m \approx k$.
\label{fig:coef}
\end{figure}

\begin{figure}[!h]
\includegraphics[width=4cm]{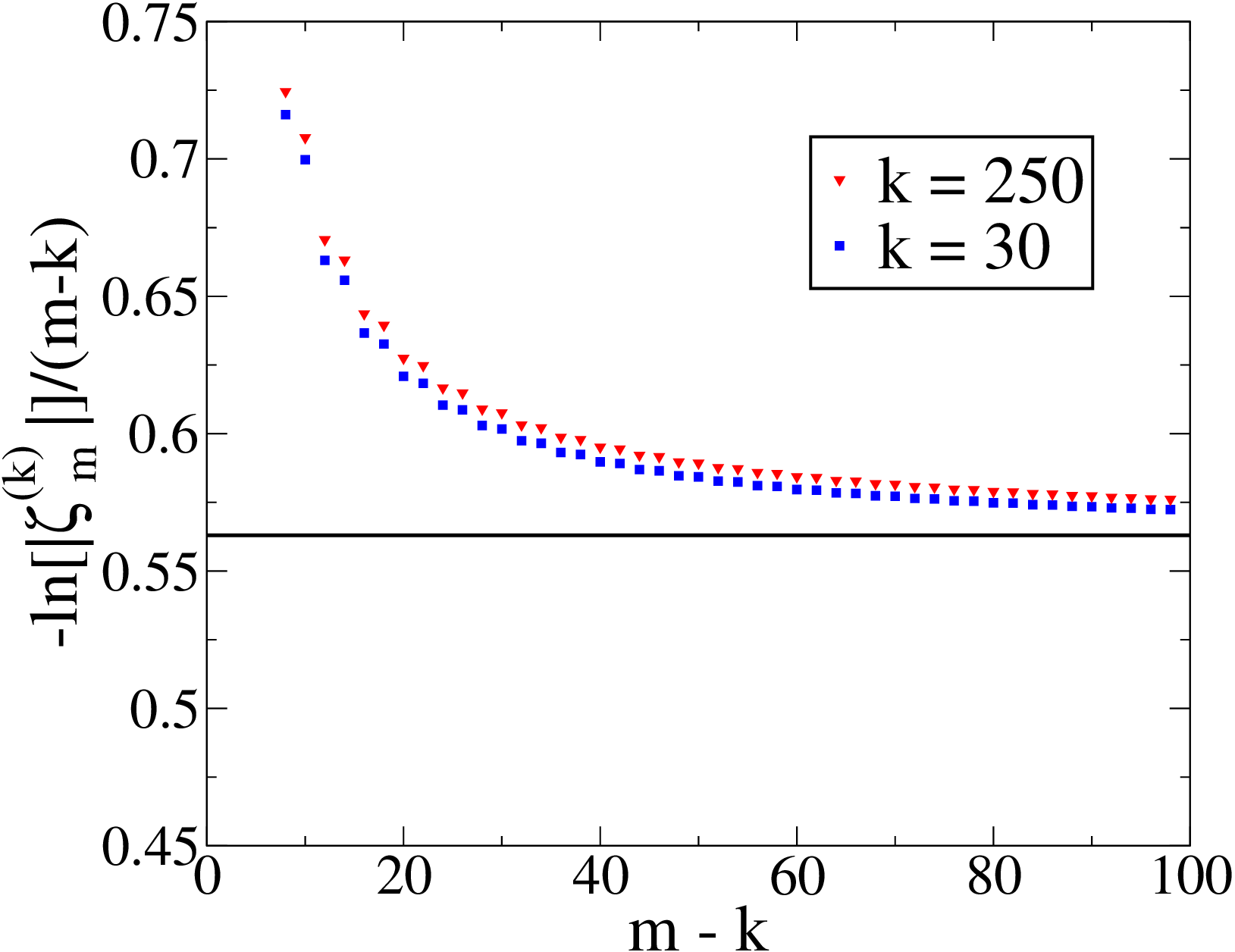}
\includegraphics[width=4cm]{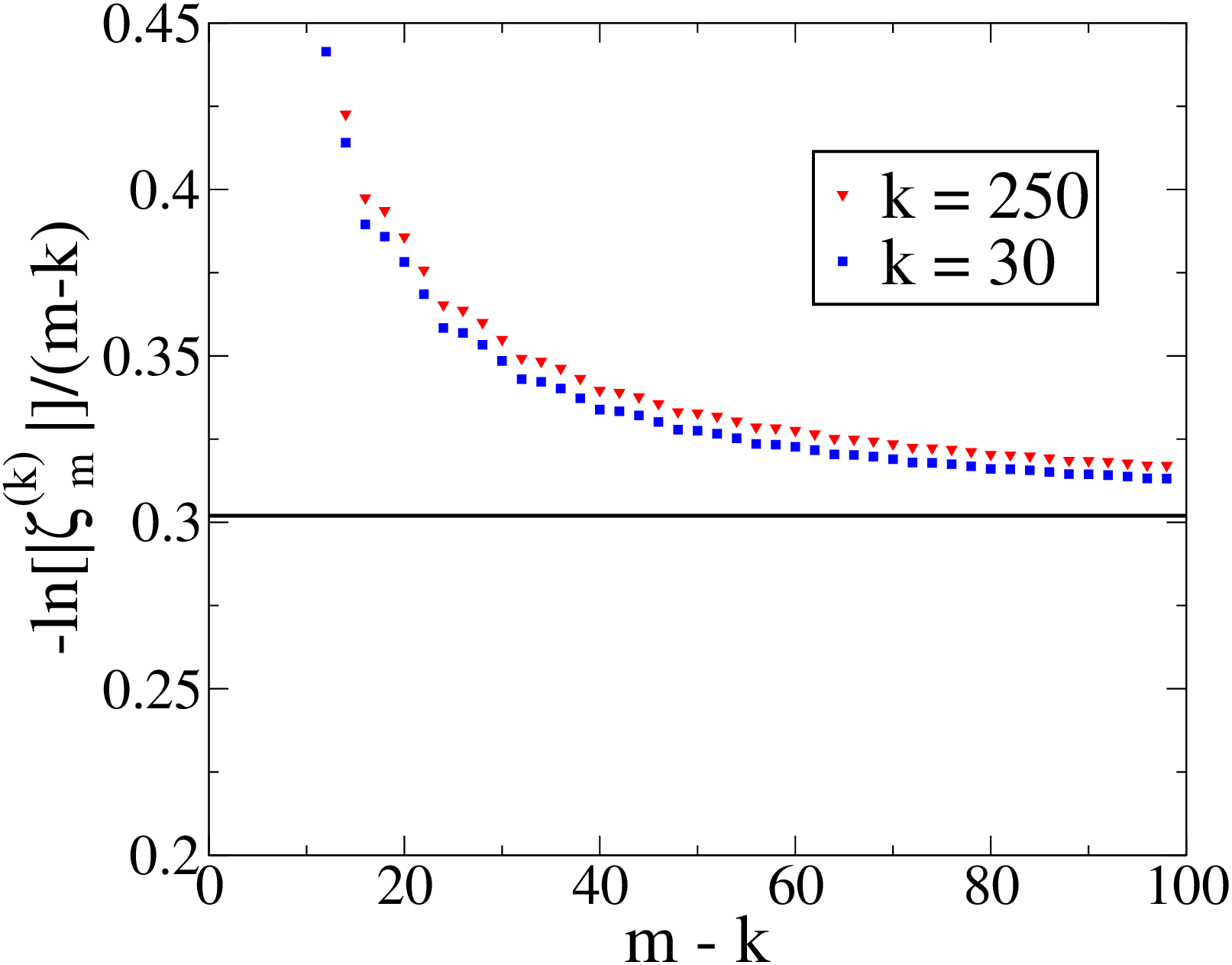}
\centering
\caption{$-\ln{(|\zeta_m^{(k)}|)}/(m-k)^2$ as function of $m-k$ for the orbital indices $k= 30,100,250$ and  $l_+/l_- = 4/5$. Left:$N=3$ and  right: $N=5$. The horizontal lines represent $\beta_N \hbar \Omega_N/(4 (N-1))$.}
\label{fig:coefr}
\end{figure}

\begin{figure}[!h]
\includegraphics[width=4cm]{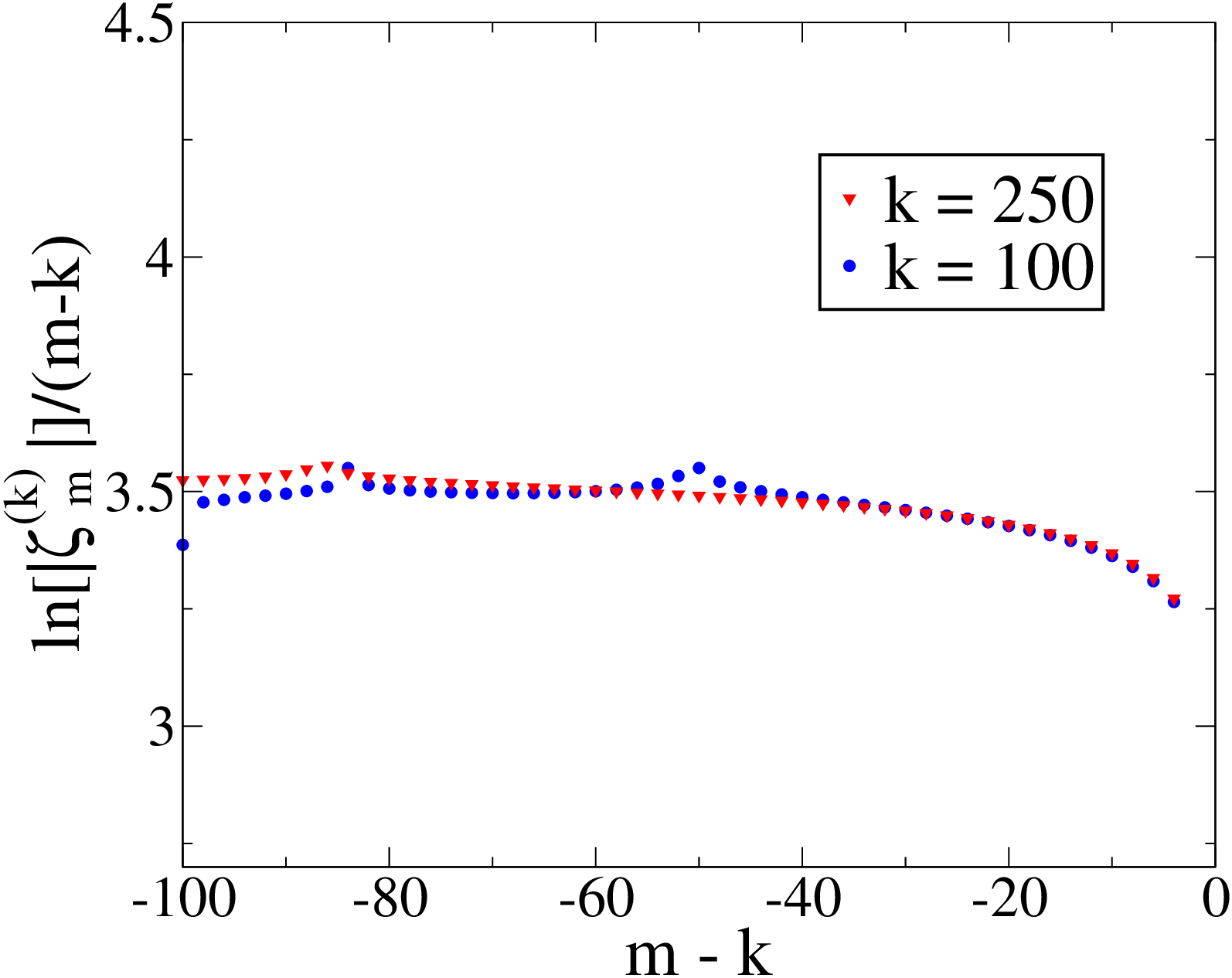}
\includegraphics[width=4cm]{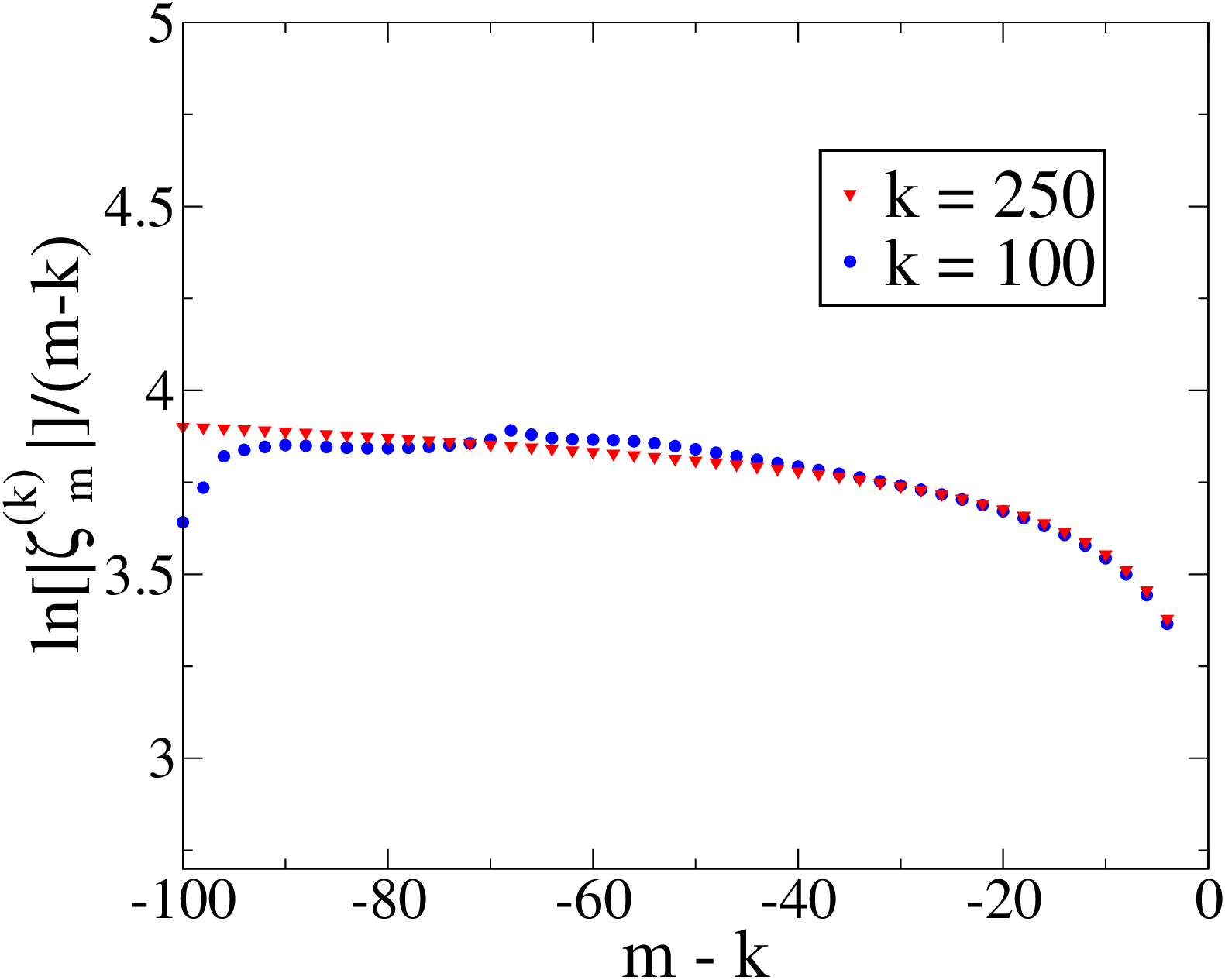}
\centering
\caption{$-\ln{(|\zeta_m^{(k)}|)}/(k-m)^2$ as function of $k-m$ for the orbital indices $k= 100,250$ and  $l_+/l_- = 4/5$. Left: $N=3$ and  right: $N=5$ }
\label{fig:coefl}
\end{figure}

One of the most remarkable results of our analysis is shown in Figure \ref{fig:coef}. Even for $l_+/l_- = 1/3$, which for $N=5$ corresponds to a very large coupling ratio $D/(m \omega^2) = 16$, the fermionic natural orbitals $\chi_k^{(f)}$ are very well approximated by a superposition of very few bosonic orbitals, $\chi_m^{(b)}$  with $m\approx k$.
To verify the Gaussian decay, Eq.\ (\ref{coefdecayrN}), for $m\gg k$ and for $m \ll k$ the exponential decay, Eq.\ (\ref{coefdecaylN}),  of the expansion coefficients $\zeta_m^{(k)} \equiv \langle \chi_m^{(b)} |\chi_k^{(f)}\rangle$   we plot  $-\ln{(|\zeta_m^{(k)}|)}/(m-k)^2$ and $-\ln{(|\zeta_m^{(k)}|)}/(k-m)^2$, respectively, as a function of $m-k$. From Figure \ref{fig:coefr}, one can infer that $\zeta_m^{(k)}$  indeed decays Gaussian-like, and the decay constants are as predicted in Eq.\ (\ref{coefdecayrN}), i.e.\ $\frac{1}{8}\beta_3 \hbar \Omega_3 \simeq 0.56$ and $\frac{1}{16}\beta_5 \hbar \Omega_5 \simeq 0.30$.
Figure \ref{fig:coefl} confirms the average exponential decay for the regime $1\ll m\ll k$.

\section{Summary and Discussion} \label{sec:Summary}
For the ground state of $N$ identical, harmonically interacting particles in a one-dimensional harmonic trap we have analytically calculated the $1$-RDO $\hat{\rho}_1$ for spinless bosons and spinless fermions in spatial representation. Usually, e.g.\ for atomic systems with Coulombic interaction, this can be done only numerically. Therefore, the result in Ref. \cite{Rob} for bosons has been extended to arbitrary $N$ and that in Ref. \cite{Dav} to arbitrary number of spinless fermions, at least in one dimension.

We have shown that $\hat{\rho}_1^{(b)}$ has the form of a Gibbs state with a Hamiltonian $\hat{H}_{eff}$ and effective temperature $\tilde{\beta}_N = \beta_N \hbar \Omega_N$. $\hat{H}_{eff}$ describes an effective harmonic oscillator with mass $M_N$, frequency $\Omega_N$ and characteristic length scale $L_N = \sqrt{\hbar/(M_N \Omega_N)}$. Consequently, for bosons the natural occupation numbers obey a Boltzmann distribution, $\lambda_k^{(b)} \sim Z_{eff}^{-1}\,e^{-\tilde{\beta}_N k}$ and the natural orbitals $\chi_k^{(b)}$ are just the Hermite functions with length scale $L_N$. For identical spinless bosons with harmonic interactions this result is expected, as pointed out at the end of Sec.\ref{sec:intro}. The advantages of the permutational invariance of the harmonic interaction (cf. Eq.\ (\ref{hamiltonianX})) is, first that $\hat{\rho}_1^{(b)}$ does not depend on the particle index and second that the parameters of $\hat{H}_{eff}$ can explicitly be determined as functions of the parameters $m, \omega$ and $D$ of the original Hamiltonian (\ref{hamiltonianX}).

For fermions, $\hat{\rho}_1^{(f)}$ contains the same Gibbs operator $e^{-\beta_N \hat{H}_{eff}}$ as well, but multiplied by a polynomial in the position operator $\hat{x}$ (cf. Eq.\ (\ref{1RDOf}) in coordinate representation or Eq.\ (\ref{1RDOfposition})). This is in contrast to free fermion models where the corresponding $\hat{\rho}_1^{(f)}$ is given by a Gibbs operator, only. The polynomial effectively results from the antisymmetry of the $N$-fermion wave function. This seems to preclude analytical calculations of the natural occupation numbers $\lambda_k^{(f)}$ and natural orbitals $\chi_k^{(f)}$. However, their asymptotic behavior for $k\rightarrow \infty$ has been derived. For fixed $N$ it is $\lambda_k^{(f)} \sim k^{N-1}\,e^{-\tilde{\beta}_N k}$, i.e.\ the fermionic character modifies the Boltzmann distribution by an additional power law factor $k^{N-1}$. Nevertheless, the dominant exponential decay is the same for bosons and fermions.

The calculation of the natural occupation numbers is in most cases performed numerically and based on a truncation of the infinite dimensional $1$-particle Hilbert spaces to a finite one. Although for identical, spinless bosons with harmonic interaction the calculation of $\hat{\rho}_1^{(b)}$ and $\lambda_k^{(b)}$ is straightforward this is not true anymore for fermions. Therefore, it seems that our results for $\lambda_k^{(f)}$ and $\chi_k^{(f)}$ are the first analytical ones for an infinite dimensional $1$-particle Hilbert space and $N>2$. The normalization of $\lambda_k^{(\alpha)}$, $\alpha= b,f$ implies $\lambda_k^{(\alpha)} \rightarrow 0$ for $k\rightarrow 0$. We have proven that this decay is exponential (for bosons it is purely exponential) and have calculated the decay constant. It would be interesting to investigate whether such an exponential decay is generic.

Although the $\lambda_k^{(\alpha)}$'s for $k\gg N$ behave very similar for bosons and fermions this is not true  anymore for the regime $k = O(N)$ or smaller. Whereas $\lambda_k^{(b)}$ exhibit a purely exponential decay for all $k$, $\lambda_k^{(f)}$ has a `discontinuity' at the `Fermi level' $k_F =N$. For zero interaction, it is
\begin{equation}\label{FermiDirac0}
\lambda_k^{(f)} = \begin{cases}
  1,  & k < k_F\\
  0, & k  \geq k_F\,.
\end{cases}
\end{equation}
With increasing interaction Fig. \ref{fig:non} demonstrates that $\lambda_k^{(f)}$ deviates from one for $k<k_F$ and from zero for $k \geq k_F$. The gap at $k_F$ becomes smaller but remains significantly large even for rather strong interactions. This behavior resembles the Fermi-Dirac distribution function. At zero ``temperature'', which corresponds to zero interaction, this distribution function is identical with the behavior in Eq.\ (\ref{FermiDirac0}). The ``softening'' of the $k$-dependence of $\lambda_k^{(f)}$ with increasing interaction strength corresponds to the softening of the Fermi-Dirac distribution for increasing temperature. It would be interesting to study $\lambda_k^{(f)}$ in the ``thermodynamic'' limit, i.e.\ $\lambda^{(f)}(\tilde{k}) = \lim_{N\rightarrow \infty} \lambda^{(f)}_{N \tilde{k}}$ and to investigate the dependence of the gap in $\lambda^{(f)}(\tilde{k})$ on the coupling constant $D$ at the Fermi level $\tilde{k}_F=1$, provided the gap survives the limit $N\rightarrow \infty$.

For bosons it is obvious from the form of $\hat{\rho}_1^{(b)}$ that the natural  orbitals are the Hermite functions, i.e.\ the eigenfunctions of the effective Hamiltonian $\hat{H}_{eff}$ (cf. \ref{Hamiltonianeff}). The results presented in Figure \ref{fig:coef} demonstrate that the fermionic natural orbitals $\chi_k^{(f)}$ are well approximated by the bosonic ones, even for the regime of strong interaction. The other relevant contributions to $\chi_k^{(f)}$ are all coming from $\chi_{k\pm 2}^{(b)},\ldots,\chi_{k\pm 2 r}^{(b)}$ with $r \ll k$, i.e.\ the natural orbitals for bosons and fermions differ only quantitatively, but not qualitatively. After all, for fixed particle number and interaction strength the similarity between both seems to become stronger with increasing $k$. Moreover, for fixed $k$ we have found a Gaussian-like decay behavior for $\langle \chi_m^{(b)}| \chi_k^{(f)}\rangle$ as function of $m-k$ in the regime $m \gg k$ (cf. Figure \ref{fig:coefr}) and an exponential one for $m\ll k$ (cf. Figure \ref{fig:coefl}), which has been derived analytically. Both decay constants do not depend on the orbital index $k$.

So far our results are valid for spinless particles. What happens if spin is also taken into account? Clearly, for bosons the new ground state is given by the original one multiplied by some spin state (which should be symmetric) and all the results from the spinless case still hold. The same is true for fermions, if additionally a sufficiently strong magnetic field is applied, which aligns all the spins parallel, along the axis of the magnetic field. In that case, the new fermionic ground state is given by the original one multiplied by the corresponding $N$-particle spin state, which is symmetric under particle exchange. Hence, all the conclusions drawn for spinless fermions still hold. However, as soon as the spin state is not symmetric anymore, the ground state in spin-orbital space is becoming more involved. Nevertheless, due to the harmonic interaction, the dominant exponential factor in $\rho_1^{(f)}(x,y)$ (cf. Eq.\ (\ref{1RDOb}) and (\ref{1RDOf})) will stay robust. Moreover, also the $1$-RDO for the excited bosonic and fermionic eigenstates are dominated by the same exponential factor and only the polynomial in front of $\rho_1^{(f/g)}$ is modified and has a higher degree.

To conclude, whereas the natural occupation numbers $\lambda_k$ for bosons and fermions differ qualitatively for $k = O(N)$ and smaller their decay behavior for $k$ large follows the same exponential dependence. The difference between the bosonic and fermionic natural orbitals is only quantitatively, even for $k = O(N)$ and smaller.

\paragraph*{Acknowledgements.---}
We thank M.\hspace{0.5mm}Christandl, D.\hspace{0.5mm}Gross, D.\hspace{0.5mm}Ebler, J.\hspace{0.5mm}Fr\"ohlich and G.M.\hspace{0.5mm}Graf for helpful discussions. We are also grateful to D.\hspace{0.5mm}Gross for bringing references \cite{Peschel1999,Peschel2003} to our attention.

We acknowledge financial support from the German Science Foundation (grant CH 843/2-1), the Swiss National Science Foundation (grants PP00P2-128455, 20CH21-138799), the Swiss National Center of Competence in Research `Quantum Science and Technology' and the Swiss State Secretariat for Education and Research supporting COST action MP1006
\\
\appendix
\section{Calculation of $\hat{\rho}_1^{(b)}(x,y)$}\label{sec:appBosons}

In the following we calculate the $1$-RDO for the bosonic ground state $\Psi_0^{(b)}$ (recall (\ref{gsbosons})):
\begin{eqnarray}
\rho_1^{(b)}(x,y)&=& \int \!\mathrm{dx_2}\ldots\mathrm{dx_N} \,\Psi_0^{(b)}(x,x_2,\ldots,x_N)^\ast \nonumber \\
&& \cdot \Psi_0^{(b)}(y,x_2,\ldots,x_N)\\
&=& \mathcal{N} ^2 e^{-(A-B_N)(x^2+y^2)} \int \!\mathrm{dx_2}\ldots\mathrm{dx_N} \,e^{-2 A (x_2^2+\ldots +x_N^2)}\nonumber \\
&& e^{2 B_N (x_2+\ldots +x_N)^2}e^{2 B_N (x+ y)(x_2+\ldots +x_N) }\nonumber
\end{eqnarray}
Here we resort to the Hubbard-Stratonovich identity,
\begin{equation}\label{hubbardstrat}
\mbox{e}^{a\xi^2}= \sqrt{\frac{a}{\pi}} \int_{-\infty}^{\infty} \!\mathrm{d}y \,\mbox{e}^{-a y^2 +2 a y \xi}
\end{equation}
for $a \in \mathbb{C}$ such that $\mbox{Re}(a)>0$.
With $a=2 B_N$ and $\xi = (x_2+\ldots + x_N)$, this leads to (for the case $B_N<0$ use a modified version of Eq.\ (\ref{hubbardstrat}) with $\xi \mapsto i \xi$)
\begin{widetext}
\begin{eqnarray}
\rho_1^{(b)}(x,y)&=&\mathcal{N} ^2 \sqrt{\frac{2 B_N}{\pi}} e^{-(A-B_N)(x^2+y^2)} \int_{-\infty}^{\infty} \!\mathrm{d}z \, \mbox{e}^{-2 B_N z^2} \int_{-\infty}^{\infty}\!\mathrm{dx_2}\ldots\mathrm{dx_N} \,e^{-2 A (x_2^2+\ldots +x_N^2)}\, e^{2 B_N (x+y+2 z)(x_2+\ldots +x_N) }\nonumber\\
&=& \mathcal{N} ^2 \sqrt{\frac{2 B_N}{\pi}} e^{-(A-B_N)(x^2+y^2)} \int_{-\infty}^{\infty} \!\mathrm{d}z \,\mbox{e}^{-2 B_N z^2} \Big(\int\!\mathrm{du} \, e^{-2 A (u-\frac{B_N}{2 A}(x+ y+2 z))^2}\Big)^{N-1}\,e^{(N-1)\frac{B_N^2}{2A} (x+ y+2 z)^2}\nonumber\\
&=&  \mathcal{N} ^2 \sqrt{\frac{2 B_N}{\pi}} \left(\frac{\pi}{2 A}\right)^{\frac{N-1}{2}}e^{-(A-B_N)(x^2+ y^2)} \int_{-\infty}^{\infty} \!\mathrm{d}z \,\mbox{e}^{-2 B_N z^2} e^{(N-1)\frac{B_N^2}{2A} (x+ y+2 z)^2} \nonumber \\
&& \nonumber
\end{eqnarray}
\end{widetext}
Since
\begin{eqnarray}
\lefteqn{\int_{-\infty}^{\infty} \!\mathrm{d}z \,\mbox{e}^{-2B_N z^2} e^{(N-1)\frac{B_N^2}{2A} (x+ y+2z)^2}}&&\nonumber \\
&=& \sqrt{\pi}\sqrt{\frac{A C_N}{(N-1)B_N^3}} e^{B_N (x+y)^2}
\end{eqnarray}
with
\begin{equation}\label{CN}
C_N = \frac{(N-1) \frac{B_N^2}{2}} {A-(N-1)B_N}
\end{equation}
we find
\begin{eqnarray}\label{reddensity}
\rho_1^{(b)}(x,y)&=& \tilde{\mathcal{N}}\,e^{-(A-B_N-C_N)(x^2+y^2)+2 C_N x y}\,,
\end{eqnarray}
where $\tilde{\mathcal{N}}$ follows from the normalization of $\rho_1^{(b)}(x,y)$.
Moreover we observe with Eqs.\ (\ref{CN}), (\ref{parameterab}) that
\begin{equation}
A-B_N-C_N =a_N\,\,,\,C_N = \frac{1}{2} b_N\,.
\end{equation}
Therefore, the exponent in Eq.\ (\ref{reddensity}) is identical to the one in Eq.\ (\ref{1RDOb}).

In Sec. \ref{sec:Bosons} we have diagonalized $\rho_1^{(b)}$ by equating it with the Gibbs state of an effective harmonic oscillator. This is equivalent to apply Mehler's formula to the expression in (\ref{reddensity}). This means to use \cite{Rob}
\begin{eqnarray}\label{Mehler}
\lefteqn{e^{-\frac{1}{4}(c^2+d^2)(z^2+\tilde{z}^2)-\frac{1}{2}(c^2-d^2) z \tilde{z}}}\nonumber \\
 &=& \sqrt{\pi}\, l (1-q^2)^{\frac{1}{2}} \sum_{k=0}^{\infty} q^k \varphi_k^{(l)}(z)\varphi_k^{(l)}(\tilde{z}) \,,
\end{eqnarray}
with $l=(c d)^{-\frac{1}{2}}$ and $q= \frac{d-c}{d+c}$. From (\ref{Mehler}) and (\ref{reddensity}) we obtain
\begin{eqnarray}
c &=&\sqrt{2(A-B_N -2 C_N)} = \sqrt{\frac{N}{\left((N-1) {l_{+}}^2+{l_{-}}^2\right)}} \nonumber \\
d &=&\sqrt{2(A-B_N)} = \sqrt{\frac{(N-1) {l_{-}}^2+{l_{+}}^2}{ N {l_{-}}^2 {l_{+}}^2}}  \nonumber \\
l &=& \sqrt{{l_{-}} {l_{+}}} \left(\frac{(N-1)l_{+}^2 + l_{-}^2}{(N-1)l_{-}^2 + l_{+}^2} \right)^{\frac{1}{4}}\,.
\end{eqnarray}
Comparing with the form in Eq.\ (\ref{1RDOb}) yields immediately the concrete expressions for the parameters $b_N$, $a_N$ and $L_N$ in Eq.\ (\ref{parameterab}).
After all the natural occupation numbers $\lambda_k^{(b)}$ (their sum is normalized to the particle number $N$) are given by
\begin{equation}\label{NON}
\lambda_k^{(b)} = N(1-q)\,q^k \,.
\end{equation}

\section{Calculation of $\rho_1^{(f)}(x,y)$}\label{sec:appFermions}
In this section we calculate the $1$-RDO $\rho_1^{(f)}(x,y)$ of the fermionic ground state $\Psi_0^{(f)}$ in spatial representation.
Below it will prove convenient to first rearrange the Vandermonde determinant
\begin{widetext}
\begin{eqnarray}
V(\vec{x})&=& \prod_{1\leq i<j\leq N} (x_i-x_j) \nonumber \\
&=& \prod_{1\leq i<j\leq N} [(x_i- s)-(x_j-s)] \nonumber \\
&=& l^{\binom{N}{2}}\,\prod_{1\leq i<j\leq N} (z_i-z_j)\qquad, z_i \equiv \frac{x_i-s}{l} \nonumber \\
&=& l^{\binom{N}{2}}\,\left|\begin{array}{lll}\,1&\ldots&\,1\\z_1&\ldots&z_N\\ \,\vdots& &\,\vdots\\ z_1^{N-1}&\ldots& z_N^{N-1} \end{array}\right|\\
&=& \left(\frac{l}{2}\right)^{\binom{N}{2}}\,\left|\begin{array}{lll}H_0(z_1)&\ldots&H_0(z_N)\\H_1(z_1)&\ldots&H_1(z_N)\\ \,\vdots& &\,\vdots\\ H_{N-1}(z_1)&\ldots&H_{N-1}(z_N) \end{array}\right|\,
\end{eqnarray}
for all $s, l \in \mathbb{C}$, where $H_k(z)$ is the $k$-th Hermite polynomial and in the last step we used the invariance of determinants under changes of a column by just linear combinations of the other ones.
Moreover, by using the orthonormalized Hermite functions $\varphi_k^{(l)}(z)$,
\begin{equation}
\varphi_k^{(l)}(z) = \frac{1}{\sqrt{2^k k!}}\,\pi^{-\frac{1}{4}}\,l^{-\frac{1}{2}}\,H_k\left(\frac{z}{l}\right) \,e^{-\frac{z^2}{2 l^2}}
\end{equation}
we find
\begin{equation}\label{VandermondeHermite}
V(\vec{x}) = const\times\left|\begin{array}{lll}\varphi_0^{(1)}(z_1)&\ldots&\varphi_0^{(1)}(z_N)\\ \varphi_1^{(1)}(z_1)&\ldots&\varphi_1^{(1)}(z_N)\\ \,\vdots& &\,\vdots\\ \varphi_{N-1}^{(1)}(z_1)&\ldots&\varphi_{N-1}^{(1)}(z_N) \end{array}\right|\,\prod_{j=1}^N \,e^{\frac{z_j^2}{2}}\,,
\end{equation}
where $z_j = z_j(x_j)$.
Note that the determinant on the rhs is nothing else but a Slater determinant.
In the following, to obtain the $1$-RDO in spatial representation we integrate out $N-1$ particle coordinates. The essential simplification used is to decouple the coordinates $x_2,\ldots,x_N$ in the exponent of the exponential function in ground state wave function (cf. Eq.\ (\ref{gsfermions})) by resorting to the Hubbard-Stratonovich identity and than afterwards using the orthogonality of the Hermite functions to make the integration trivial. In order not to confuse the reader we do not care about global constants, collect and represent them just by symbols $\mathcal{N}^{(i)},i=1,,\ldots$ and normalize the final expression for the $1$-RDO at the end.
We find
\begin{eqnarray}
\rho_1^{(f)}(x,y)&=& \int \!\mathrm{d}x_2\ldots \mathrm{d}x_N \, \Psi_N(x,x_2,\ldots,x_N)^\ast \Psi_N(y,x_2,\ldots,x_N) \nonumber \\
&=& \mathcal{N}^{(1)} \, e^{-(A-B_N)(x^2+y^2)} \, \int \!\mathrm{d}x_2\ldots \mathrm{d}x_N \, V(x,x_2,\ldots,x_N) V(y,x_2,\ldots,x_N) \nonumber \\
&&\cdot e^{-2A (x_2^2+\ldots+x_N^2)} \, e^{2B_N (x_2+\ldots+x_N)^2}\,e^{2B_N (x+y)(x_2+\ldots+x_N)}\,.
\end{eqnarray}
Now we use the Hubbard-Stratonovich identity (\ref{hubbardstrat}) with
\begin{equation}
a\equiv 2B_N\qquad,\, \xi\equiv x_2+\ldots +x_N
\end{equation}
to decouple the mixed terms in the exponent $(x_2+\ldots +x_N)^2$. This yields
\begin{eqnarray}
\rho_1^{(f)}(x,y)&=& \mathcal{N}^{(2)}\,e^{-(A-B_N)(x^2+y^2)}\, \int \!\mathrm{d}z \, \int \!\mathrm{d}x_2\ldots \mathrm{d}x_N \, V(x,x_2,\ldots,x_N) V(y,x_2,\ldots,x_N)\,e^{-2A (x_2^2+\ldots+x_N^2)} \nonumber \\
&&\cdot e^{2B_N (x+y)(x_2+\ldots+x_N)} \, e^{-2B_N z^2}\, e^{4B_N(x_2+\ldots+x_N)z}\nonumber \\
&=& \mathcal{N}^{(2)}\,e^{-(A-B_N)(x^2+y^2)}\, \int \!\mathrm{d}z \, e^{-2B_N z^2} \int \!\mathrm{d}x_2\ldots \mathrm{d}x_N \, V(x,x_2,\ldots,x_N) V(y,x_2,\ldots,x_N) \nonumber \\
&& \cdot\prod_{j=2}^N \,e^{-2A x_j^2  + \left(2B_N (x+y)+4B_N z\right) x_j} \nonumber \\
&=& \mathcal{N}^{(2)}\,e^{-(A-B_N)(x^2+y^2)}\, \int \!\mathrm{d}z \, e^{-2B_N z^2} \int \!\mathrm{d}x_2\ldots \mathrm{d}x_N \, V(x,x_2,\ldots,x_N) V(y,x_2,\ldots,x_N) \nonumber \\
&& \cdot\prod_{j=2}^N \,e^{-2A \big(x_j - \frac{B_N}{2A} (x+y+2z)\big)^2} \, e^{\frac{B_N^2}{2A} (x+y+2z)^2} \,.
\end{eqnarray}
Now we fix $s$ introduced above. For $\,j=2,3,\ldots,N$ we use
\begin{equation}\label{Zvariables}
z_j \equiv \frac{x_j-s}{l} = \sqrt{2A} \,\big(x_j - \frac{B_N}{2A} (x+y+2z)\big)
\end{equation}
with
\begin{equation}
l \equiv \frac{1}{\sqrt{2A}} \qquad,\, s \equiv  \frac{B_N}{2A}(x+y+2z) \,.
\end{equation}
Thus, by using (\ref{VandermondeHermite}) and $z_1^{(X)}\equiv \left(\frac{x-s}{l}\right)$, $z_1^{(Y)}\equiv \left(\frac{y-s}{l}\right)$,  we find
\begin{eqnarray}\label{Slatertrick}
\rho_1^{(f)}(x,y)&=& \mathcal{N}^{(3)}\,e^{-(A-B_N)(x^2+y^2)}\, \int \!\mathrm{d}z \, e^{-2B_N z^2} \,e^{\frac{B_N^2}{2A}(N-1)(x+y+2z)^2} \,e^{\frac{\left(z_1^{(X)}\right)^2+\left(z_1^{(Y)}\right)^2}{2}} \\
&& \cdot \int \!\mathrm{d}z_2\ldots \mathrm{d}z_N   \left|\begin{array}{llll}\varphi_0^{(1)}\left(z_1^{(X)}\right)&\varphi_0^{(1)}(z_2)&\ldots&\varphi_0^{(1)}(z_N)\\ \varphi_1^{(1)}\left(z_1^{(X)}\right)&\varphi_1^{(1)}(z_2)&\ldots&\varphi_1^{(1)}(z_N)\\ \,\vdots&& &\,\vdots\\ \varphi_{N-1}^{(1)}\left(z_1^{(X)}\right)&\varphi_{N-1}^{(1)}(z_2)&\ldots&\varphi_{N-1}^{(1)}(z_N) \end{array}\right| \, \left|\begin{array}{llll}\varphi_0^{(1)}\left(z_1^{(Y)}\right)&\varphi_0^{(1)}(z_2)&\ldots&\varphi_0^{(1)}(z_N)\\ \varphi_1^{(1)}\left(z_1^{(Y)}\right)&\varphi_1^{(1)}(z_2)&\ldots&\varphi_1^{(1)}(z_N)\\ \,\vdots&& &\,\vdots\\ \varphi_{N-1}^{(1)}\left(z_1^{(Y)}\right)&\varphi_{N-1}^{(1)}(z_2)&\ldots&\varphi_{N-1}^{(1)}(z_N) \end{array}\right| \nonumber\,.
\end{eqnarray}
The orthogonality of the Hermite functions makes the $z_2,\ldots,z_N$ integrals trivial and we find
\begin{equation}
\rho_1^{(f)}(x,y)= \mathcal{N}^{(4)}\,e^{-(A-B_N)(x^2+y^2)}\, \int \!\mathrm{d}z \, e^{-2B_N z^2} \,e^{\frac{B_N^2}{2A}(N-1)(x+y+2z)^2}\, \sum_{k=0}^{N-1}\,\frac{1}{2^k k!}\, H_k\left(z_1^{(X)}\right)  H_k\left(z_1^{(Y)}\right)\,.
\end{equation}
Finally, we simplify the $z$-integral.
We rearrange
\begin{eqnarray}
\lefteqn{2B_N z^2- \frac{B_N^2}{2A} (N-1)(x+y+2z)^2} \nonumber \\
&=& \left(2 B_N - \frac{2B_N^2}{A}(N-1)\right)\,z^2 - 2 \frac{B_N^2}{A}(N-1)(x+y)\,z- \frac{B_N^2}{2A}(N-1)(x+y)^2 \nonumber \\
&\equiv& r \, z^2- 2t\,z+v\nonumber \\
&=& r\,\left(z-\frac{t}{r}\right)^2-\frac{t^2}{r}+v
\end{eqnarray}
with
\begin{equation}\label{parameterr}
r\equiv 2B_N\,\left(1 - \frac{B_N}{A}(N-1)\right)\,\,,\, t\equiv \frac{B_N^2}{A}(N-1)(x+y)\,\,,\, v\equiv - \frac{B_N^2}{2A}(N-1)(x+y)^2\,.
\end{equation}
From Eq.\ (\ref{parameterr}) it follows with Eq.\ (\ref{CN})
\begin{eqnarray}
\frac{t^2}{r}-v &=& \frac{B_N^3(N-1)^2}{2A \left(A-B_N(N-1)\right)}\,(x+y)^2+\frac{B_N^2}{2A}\,(x+y)^2\nonumber \\
&=& C_N\,(x+y)^2 \nonumber \\
\frac{t}{r} &=& \frac{B_N(N-1)}{2\left(A-B_N(N-1)\right)}\,(x+y) = \frac{C_N}{B_N}\,(x+y)
\end{eqnarray}
and we obtain
\begin{eqnarray}\label{1RDOfUint}
\rho_1^{(f)}(x,y)&=& \mathcal{N}^{(5)}\,e^{-\left(A-B_N-C_N\right)\,(x^2+y^2)+ 2 C_N\,x y} \nonumber \\
&&\cdot  \int \!\mathrm{d}u \, e^{- u^2}\,\sum_{k=0}^{N-1}\,\frac{1}{2^k k!}\, H_k(p u+q(x,y))  H_k(p u+q(y,x))\,,
\end{eqnarray}
where we defined
\begin{equation}
p\equiv \sqrt{\frac{B_N}{A-B_N(N-1)}}\qquad,\,q(x,y)= \sqrt{2A}\left[x-\frac{B_N}{2\left(A-B_N(N-1)\right)}\,(x+y)\right]\,.
\end{equation}
Note that the exponential factor in Eq.\ (\ref{1RDOfUint}) is identical to the corresponding factor in Eq.\ (\ref{reddensity}) for $\rho_1^{(b)}(x,y)$.
From the fact that only even order terms in $u$ are relevant for $u$-integration in Eq.\ (\ref{1RDOfUint}) and due to the structure of the Hermite polynomials it is clear that the $1$-RDO has the form
\begin{equation}
\rho_1^{(f)}(x,y) = F_N(x,y)\, \exp{\left[-a_N (x^2+y^2) +b_N x y\right]},
\end{equation}
with
\begin{equation}
F_N(x,y) = \sum_{\nu=0}^{N-1} \sum_{\mu=0}^{2 \nu} \,c_{\nu,\mu}\, x^{2\nu-\mu} y^{\mu}\,.
\end{equation}
The coefficients $c_{\nu,\mu}$ depend on the model parameters and fulfill $c_{\nu,\mu} = c_{\nu,2\nu-\mu}$ and $a_N, b_N$ are given by Eq.\ (\ref{parameterab}).
\\
\end{widetext}

\section{Eigenvalue Equation for the Fermionic Matrix $(\langle \varphi_m|\hat{\rho}_1^{(f)}|\varphi_n\rangle)$}\label{sec:appEigenvalue}
With $\hat{x}$ the position operator and recalling the representation $\rho_1^{(f)}(x,y) = \langle x| e^{-\beta_N \hat{H}_{eff}}|y\rangle$ we get from Eq.\ (\ref{1RDOf})
\begin{equation}\label{1RDOfposition}
\hat{\rho}_1^{(f)} = \sum_{\nu=0}^{N-1} \sum_{\mu=0}^{2\nu} \,c_{\nu,\mu}\,\hat{x}^{2\nu-\mu}\, e^{-\beta_N \hat{H}_{eff}} \,\hat{x}^{\mu},
\end{equation}
which is hermitian due to $c_{\nu,\mu} = c_{\nu,2\nu-\mu}$. Since $\hat{H}_{eff}$ describes a harmonic oscillator with characteristic length scale $L_N$ (see Sec. \ref{sec:Bosons}) $\hat{x}$ and $\hat{H}_{eff}$ can elegantly be expressed by the corresponding creation and annihilation operators
\begin{eqnarray}
\hat{x}&=& \sqrt{\frac{L_N}{2}}\,(a+a^\dagger)\nonumber \\
\hat{H}_{eff} &=& \hbar \Omega_N (a^\dagger a + \frac{1}{2})\,.
\end{eqnarray}
Then, $\hat{\rho}_1^{(f)}$ takes the form
\begin{eqnarray}\label{1RDOfladderop}
\hat{\rho}_1^{(f)} &=& \sum_{\nu=0}^{N-1}\left(\frac{L_N}{2}\right)^{\nu} \sum_{\mu=0}^{2\nu} \,c_{\nu,\mu}\,(a+a^\dagger)^{2\nu-\mu}\nonumber \\
&&\cdot e^{-\beta_N \hbar \Omega_N (a^\dagger a +\frac{1}{2})} \,(a+a^\dagger)^{\mu}\,.
\end{eqnarray}
To determine the natural orbitals $|\chi^{(f)}\rangle$ of $\hat{\rho}_1^{(f)}$ we expand them w.r.t. the bosonic natural orbitals, the Hermite states $|m\rangle$ with natural length scale $L_N$ ($\varphi_m^{(L_N)}(x)\equiv \langle x |m\rangle$):
\begin{equation}
|\chi^{(f)}\rangle = \sum_{m=0}^{\infty}\,\zeta_m\,|m\rangle\,.
\end{equation}
Since $a^\dagger a |m\rangle = m |m\rangle$ we find for $\mu$ fixed and $m$ sufficiently large
\begin{equation}
(a+a^\dagger)^{\mu}|m\rangle = m^{\frac{\mu}{2}}\,\left(1+O\left(\frac{1}{m}\right)\right) \sum_{\kappa=0}^{\mu}\binom{\mu}{\kappa}|m-\mu-\kappa\rangle\,.
\end{equation}
Using this asymptotic result we get for $N$ fixed and $m\rightarrow \infty$
\begin{widetext}
\begin{eqnarray}\label{coefrelNOf}
\hat{\rho}_1^{(f)}|m\rangle &\rightarrow& m^{N-1} e^{-\beta_N\hbar \Omega_N (m+\frac{1}{2})}\sum_{\nu=0}^{N-1}\left(\frac{L_N}{2}\right)^{\nu}\sum_{\mu=0}^{2 \nu} c_{\nu,\mu} \sum_{\kappa=0}^{\mu} \binom{\mu}{\kappa} e^{\beta_N \hbar \Omega_N (\mu-2\kappa)}\sum_{\tau=0}^{2\nu-\mu} \binom{2\nu-\mu}{\tau} |m-2\underbrace{(\nu-\kappa-\tau)}_{:=r}\rangle \nonumber \\
&=& m^{N-1} e^{-\beta_N\hbar \Omega_N (m+\frac{1}{2})} \sum_{r=-(N-1)}^{N-1} h_{m,m-2r} |m-2r\rangle\,,
\end{eqnarray}
\end{widetext}
where the real coefficients $h_{m,m-2r}$ depend on $L_N$ and $\beta_N \hbar \Omega_N$, but not \emph{explicitly} on $m$.

\bibliography{refNO,comments}

\end{document}